\documentclass{article}
\usepackage[preprint]{spconf}
\usepackage{amsmath,graphicx,url,times,amsmath,amssymb,acronym,graphicx,balance}
\usepackage{color,booktabs}
\usepackage[table]{xcolor}
\usepackage{cite}
\usepackage{lipsum}

\copyrightnotice{\copyright\ IEEE 2021}

\title{Towards efficient models for real-time deep noise suppression}
\name{Sebastian Braun, Hannes Gamper, Chandan K.A. Reddy, Ivan Tashev} % \textbf{}
\address{
	Microsoft Research, Redmond, WA, USA \\
	sebastian.braun@microsoft.com}

\acrodef{STFT}{short-time Fourier transform}
\acrodef{MSE}{mean-squared error}
\acrodef{MAE}{mean absolute error}
\acrodef{PSD}{power spectral density}
\acrodef{RTF}{relative transfer function}
\acrodef{SNR}{signal-to-noise ratio}
\acrodef{PDF}{probability density function}
\acrodef{DOA}{direction-of-arrival}
\acrodef{VAD}{voice activity detector}
\acrodef{MVDR}{minimum variance distortionless response}
\acrodef{AIR}{acoustic impulse response}
\acrodef{PESQ}{perceptual evaluation of speech quality}
\acrodef{STOI}{short-time objective intelligibility}
\acrodef{LSD}{log spectral distance}
\acrodef{CD}{cepstral distance}
\acrodef{WER}{word error rate}
\acrodef{SPP}{speech presence probability}
\acrodef{DNN}{deep neural network}
\acrodef{RNN}{recurrent neural network}
\acrodef{CNN}{convolutional neural network}
\acrodef{FC}{fully connected}
\acrodef{CRN}{convolutional recurrent network}
\acrodef{LSTM}{long-term short-term}
\acrodef{GRU}{gated recurrent unit}
\acrodef{FF}{feed forward}
\acrodef{ReLU}{rectified linear unit}
\acrodef{GCC}{generalized cross-correlation}
\acrodef{RMSE}{root-mean-square error}
\acrodef{CPSD}{cross-power spectral density}
\acrodef{siSDR}{scale-invariant signal-to-distortion ratio}
\acrodef{SDR}{signal-to-distortion ratio}
\acrodef{MagMSE}{Magnitude MSE}
\acrodef{LPS}{logarithmic power spectrum}
\acrodef{CSE}{complex spectrum error}
\acrodef{MagSE}{magnitude spectrum error}
\acrodef{LogMagSE}{logarithmic magnitude spectrum error}
\acrodef{DL}{deep learning}
\acrodef{PLSD}{phase-aware logarithmic spectrum distance }
\acrodef{SDW}{speech distortion-weighted}
\acrodef{MOS}{mean opinion score}
\acrodef{RIR}{room impulse response}
\acrodef{MAC}{multiply-accumulate}
\acrodef{CRUSE}{Convolutional Recurrent U-net for Speech Enhancement}
\acrodef{convGLU}{convolutional gated linear unit}
\acrodef{FD}{frequency-domain}

\definecolor{matlab1}{rgb}{0, 0.4470, 0.7410}
\definecolor{matlab2}{rgb}{0.8500, 0.3250, 0.0980} 
\definecolor{matlab3}{rgb}{0.9290, 0.6940, 0.1250} 
\definecolor{matlab4}{rgb}{0.4940, 0.1840, 0.5560} 
\definecolor{matlab5}{rgb}{0.4660, 0.6740, 0.1880}

\begin{document}
	\ninept
	
	\maketitle
	\begin{abstract}
		With recent research advancements, deep learning models are becoming attractive and powerful choices for speech enhancement in real-time applications. While state-of-the-art models can achieve outstanding results in terms of speech quality and background noise reduction, the main challenge is to obtain compact enough models, which are resource efficient during inference time.
		An important but often neglected aspect for data-driven methods is that results can be only convincing when tested on real-world data and evaluated with useful metrics. 
		In this work, we investigate reasonably small recurrent and convolutional-recurrent network architectures for speech enhancement, trained on a large dataset considering also reverberation. We show interesting tradeoffs between computational complexity and the achievable speech quality, measured on real recordings using a highly accurate MOS estimator. It is shown that the achievable speech quality is a function of network complexity, and show which models have better tradeoffs.
	\end{abstract}
	

	\section{Introduction}
\label{sec:intro}
%\lipsum
Speech enhancement using neural networks has seen large attention in research in the recent years \cite{Wang2018} and is starting to be deployed in commercial human-to-human communication applications. While the trend in research still majorly follows the trajectory of developing larger networks to further improve the performance and quality, for real-world applications following the opposite trend is of much higher interest: \emph{How to obtain the best speech quality given a maximum computational budget?} Running current state-of-the-art noise suppression neural networks is still challenging on resource limited devices, where noise suppression is often only a small fraction among several other tasks running on the devices, such as other audio processing tasks, video, encoding, transmission, etc.

Earlier network architectures were mainly \ac{RNN} structures, which were believed promising in terms of efficiency due to its efficient temporal modeling capabilities \cite{Weninger2015,Williamson2017,Xia2020}. While such models seem to hit a performance saturation, the use of \acp{CRN} and \acp{CNN} raised the performance, but resulted in development of enormously large architectures \cite{Wichern2017,Strake2019,Tan2018,Wisdom2019} that are impractical to run on typical edge devices like consumer laptops, mobile phones, or even less powerful devices like wearables or hearing aids.
Efficient models are also obtained by building as much prior knowledge into the models as possible, rather than trying to learn well-understood blocks such as time-frequency transforms from scratch. While time-domain networks such as \cite{Luo2019} could in theory yield superior performance than \ac{FD} networks, proof of generalization on real data in reverberant environments and real recordings has not yet been shown \cite{Maciejewski2020}. Therefore, we stick in this work to \ac{FD} implementations.

To draw valid and general conclusions from our experiments, we train on large scale data simulating the most important aspects of real-world data such as reverberation, many different speakers, a vast amount of noise types, and varying microphone signal levels. We propose a powerful data generation and augmentation pipeline that deals with reverberant and non-reverberant speech to reduce heavy reverberation, using signal-based estimates of reverberation parameters.
Results are shown on real recordings of a public dataset using a \ac{DNN} based MOS predictor that has shown high correlation to subjective ratings in practice.

In this work, we compare \ac{RNN} with \ac{CRN} architectures and show which network parts can be scaled, removed, or replaced by more efficient modules, at which gains in complexity and which loss in quality. Specifically, we investigate the influence of \ac{RNN} size, type, and the use of disconnected parallel \acp{RNN}. For \acp{CRN} with a symmetric convolutional encoder/decoder structure, we investigate the convolution layers, spectral vs. spectro-temporal convolutions, and skip connections.
As a result, we propose an efficient \ac{CRN} structure with around 4-5 times less computational operations with similar quality than previously proposed \acp{CRN}.

%
% \cite{Williamson2017,Ephrat2018,Luo2019}, or computationally expensive network architectures \cite{Williamson2017,Strake2019,Wichern2017,Wisdom2019,Tan2018}. 
%
%of enhancement filters or masks \cite{Wang2014,Williamson2017} to signal-based metrics, such as spectral magnitude-based \ac{MSE}, phase-sensitive \ac{MSE} \cite{Kolbaek2017} and finally the complex spectral \ac{MSE} \cite{Strake2019}.
%
%The study in \cite{Kolbaek2020} 

%\section{Signal Model}
%\label{sec:sigmodel}

\section{Enhancement system and training objective}
\label{sec:enhancement_system}
\begin{figure}[b]
	\centering
	\includegraphics[width=\columnwidth,clip,trim=180 160 190 190]{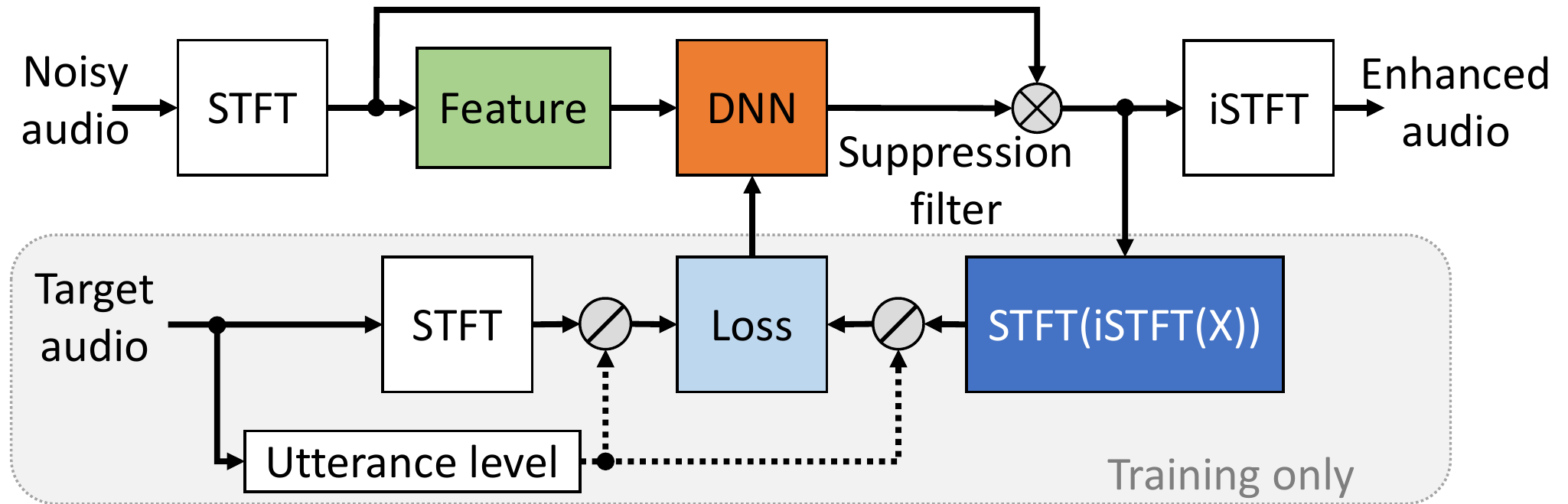}
	\caption{Enhancement system and training with STFT consistency \cite{Wisdom2019} and level-invariant loss \cite{Braun2020}.}
	\label{fig:enhancement_system}
\end{figure}

We use spectral suppression-based enhancement systems due to their robust generalization, logical interpretation and control, and easier integration with existing speech processing algorithms. The input features to the networks are log power spectra. The network predicts a real-valued, time-varying suppression gain per time-frequency bin, that is applied to the complex input spectrum, and transformed back to time-domain as shown in Fig.~\ref{fig:enhancement_system} in the upper branch.
To compute a single frame, the network requires only the feature of the current frame, or when using causal convolutions, also several past frames. Therefore, the algorithmic delay of the systems depend only on the \ac{STFT} window size. 

We train the networks enforcing \ac{STFT} consistency \cite{Wisdom2019} by propagating the \ac{FD} output through reconstruction and another \ac{STFT} to compute a \ac{FD} loss as shown in Fig.~\ref{fig:enhancement_system}. This preserves the flexibility of integrating the network with other \ac{FD} algorithms, and offloading the \ac{STFT} operations to optimized implementations. 
As shown in Fig.~\ref{fig:enhancement_system}, each training sequence, i.\,e.\, predicted and target signals, are normalized by the active target utterance level, to ensure balanced optimization for signal-level dependent losses \cite{Braun2020}.

We train on the complex compressed \ac{MSE} loss \cite{Ephrat2018}, blending the magnitude-only with a phase-aware term, which we found to be superior to other losses for reverberant speech enhancement \cite{Braun2020a}. The loss function per sequence is given by
\begin{equation}
	\label{eq:loss}
	\mathcal{L} = (1\!-\!\lambda) \sum_{k,n} \left||S|^c \!-\! |\widehat{S}|^c\right|^2 + \lambda \sum_{k,n} \left||S|^c e^{j\varphi_{S}} \!-\! |\widehat{S}|^c e^{j\varphi_{\widehat{S}}} \right|^2,
\end{equation}
where $c=0.3$ is a compression factor, $\lambda=0.3$ \cite{Braun2020a} is a weighting between complex and magnitude-based loss, and we omitted the dependency of the target speech spectral bins $S(k,n)$ on the frequency and time indices $k,n$ for brevity. 

The networks are trained in batches of 10 sequences of 10~s length using the AdamW optimizer \cite{Loshchilov2019}, learning rate of $8 \cdot 10^{-5}$, and weight decay of 0.1. The best model is picked based on the validation metric using a heuristic optimization criterion $Q$ using \ac{PESQ} \cite{ITU_T_P862}, \ac{siSDR} \cite{Roux2019} and \ac{CD} \cite{Kitawaki1988}:
\begin{equation}
	Q = PESQ + 0.2\cdot siSDR - CD.
\end{equation}

\section{Network architectures}
In this section, we describe \ac{RNN} and \ac{CRN} architectures and modify them to improve efficiency. All models use the same features, prediction targets, loss, and training strategy described in Section~\ref{sec:enhancement_system}.
\begin{figure}[tb]
	\centering
	\includegraphics[width=.9\columnwidth,clip,trim=300 178 290 158]{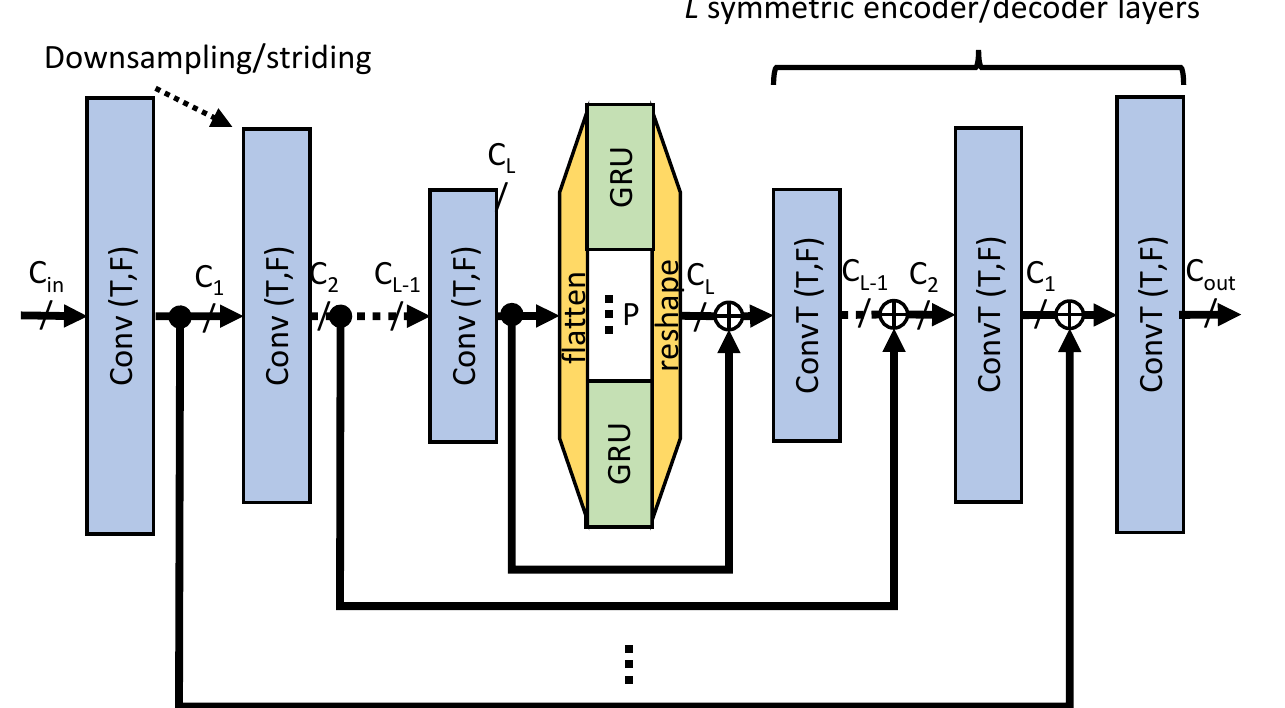}
	\caption{CRUSE network architecture with $L$ encoder/decoder layers and a bottleneck with $P$ parallel recurrent layers.}
	\label{fig:architectures}
	\vspace{-.3cm}
\end{figure}

\subsection{NSnet2}
The network proposed in \cite{Braun2020}, referred to as \emph{NSnet2}, consists only of \ac{FC} and \ac{GRU} \cite{Cho2014} layers in the format FC-GRU-GRU-FC-FC-FC. All \ac{FC} layers are followed by \ac{ReLU} activations, except the last layer has sigmoid activations to predict a constrained suppression gain. The standard layer dimensions are 400 for \acp{GRU}, 600 for \ac{FC} layers, i.\,e.\, 400-400-400-600-600-$K$, but we also investigate different configurations. The input and output dimensions are the number of frequency bins $K$.

\subsection{CRUSE}
The second network is a \ac{CRN} U-Net structure derived from \cite{Tan2018}, referred to in the remainder as \emph{\ac{CRUSE}}. As shown in Fig.~\ref{fig:architectures}, the structure has $L$ symmetric convolutional and deconvolutional encoder and decoder layers with kernels of size $(2,3)$ in time and frequency dimensions. The convolution kernels move with a stride of $(1,2)$, i.\,e.\ downsample the features along the frequency axis efficiently, while the number of channels $C_\ell$ for layer $\ell=\{1,\hdots,L\}$ increase per encoder layer, and decrease mirrored in the decoder. In this work, input and output channels $C_\text{in} = C_\text{out} = 1$, but they can be extended to e.\,g.\, take complex values or multiple features as multiple channels. Convolutional layers are followed by leaky \ac{ReLU} activations, while the last layer uses sigmoid. Between encoder and decoder sits a recurrent layer, which is fed with all features flattened along the channels. In \cite{Tan2018} a stack of two \ac{LSTM} layers was proposed at this stage. As will be shown in our experimental results in Section~\ref{sec:results}, replacement by a single \ac{GRU} layer yields very little performance loss, but huge computational savings. A \ac{GRU} saves 25\% complexity compared to an \ac{LSTM} layer.
Two further modifications are addressed in the following two paragraphs.

\textbf{Parallel RNN grouping}
As will be shown in Section~\ref{sec:results}, the performance of both CRUSE and NSnet2 directly scales with the bottleneck size, i.\,e.\ the width $R$ of the central \ac{RNN} layer(s). However, the complexity of \ac{RNN} layers scales with $R^2$, making wide \acp{RNN} computationally unattractive. Therefore, we adopt the technique proposed in \cite{Tan2020b}, to group the wide fully connected \acp{RNN} into $P$ disconnected parallel \acp{RNN}, still yielding the same forward information flow as shown in Fig.~\ref{fig:architectures}.
We denote the number of $P$ parallel \acp{GRU}, where $P\!=\!1$ means the last convolutional encoder output is flattened to a single vector and fed to a single \ac{GRU}, while with $P\!>\!1$, the encoder output is reshaped to $P$ vectors of same length, being fed through $P$ disconnected \acp{GRU}, and being reshaped again to the number of decoder channels $C_L$.
Another practical advantage is the possible parallel execution of the disconnected \acp{RNN}.

\begin{figure}[tb]
	\centering
	\includegraphics[width=.85\columnwidth,clip,trim=220 202 210 160]{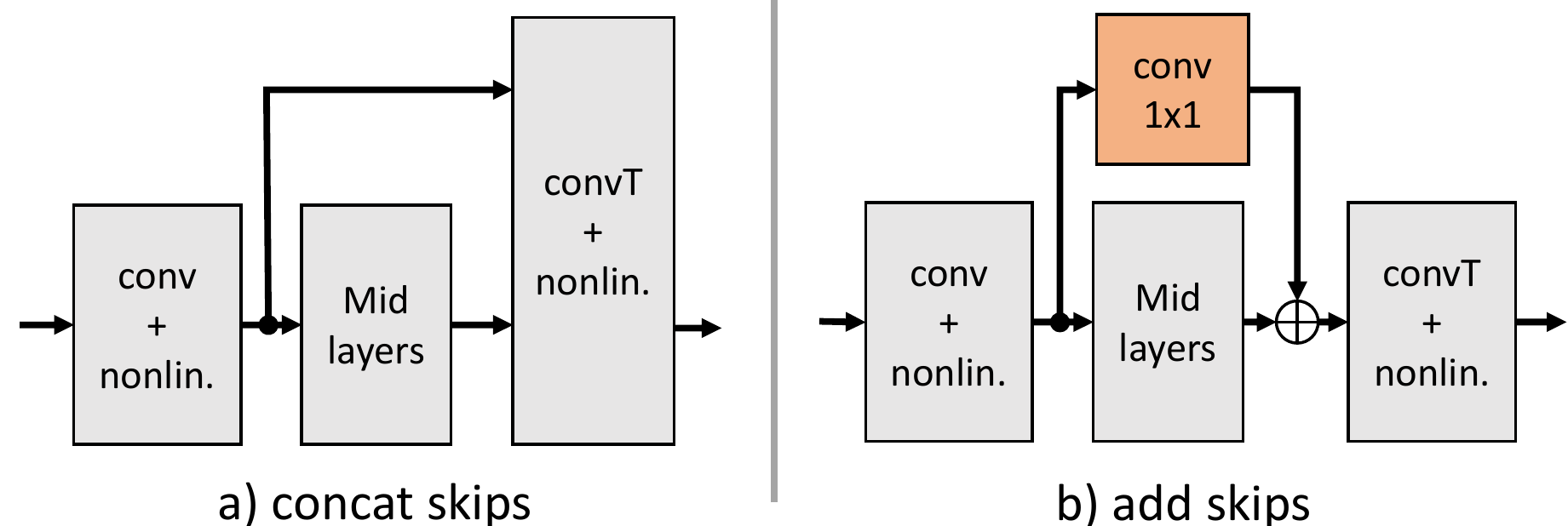}
	\caption{Skip connections by a) doubling the decoder input channels, b) addition. We found inserting $1\!\!\times\!\!1$ convolutions in the add skips connections useful.}
	\label{fig:skips}
	\vspace{-.2cm}
\end{figure}

\textbf{Skip connections}
As shown in Fig.~\ref{fig:architectures}, each convolutional encoder layer is connected to its corresponding decoder layer by a skip connection. In \cite{Tan2020b} skip connections between corresponding encoder and decoder layers have been implemented by concatenating the encoder output to the corresponding decoder input along the channel dimension as shown in Fig.~\ref{fig:skips}a). 
This doubles the number of decoder input channels, resulting in higher complexity. More efficient skip connections are implemented by simply adding the encoder outputs to the decoder inputs, resulting in minor performance degradation. We found when adding a trainable channel-wise scaling and bias in the add-skip connections, which can be implemented as convolutions with $C_\ell$ channels and (1,1) kernels as in Fig.~\ref{fig:skips}b), therefore being very cheap, improves the performance at negligible additional cost.

%\subsubsection{Fancy convolutions}
%Lastly, we also explore replacing the standard convolution operations in encoder/decoder in Fig.~\ref{fig:architectures} by supposedly more complicated convolution operations: i) Depthwise separable convolutions, which have been shown in some tasks to be less complex with little performance loss.
%ii) \acp{convGLU} have been shown to improve performance over standard convolutional layers e.\,g.\, in ConvTASnet \cite{Luo2019}. A big caveat is that a \acp{convGLU} layer exactly doubles the computational burden of a conventional layer. The gained performance advantage has therefore to be evaluated against the increased complexity.

\section{Experimental setup}

\subsection{Dataset}
We a use large-scale synthetic training set and test on real recordings to ensure generalization of our results to real-world signals. The training set uses 544~h of high \ac{MOS} rated speech recordings from the LibriVox corpus, 247~h noise recordings from Audioset, Freesound, internal noise recordings and 1~h of colored stationary noise. Except for the 65~h internal noise recordings, the data is available publicly as part of the 2nd DNS challenge\footnote{https://github.com/microsoft/DNS-Challenge/tree/icassp2021-final}.
We estimated $T_{60}$ and $C_{50}$ for each speech file using \cite{Gamper2018,Gamper2020} and classified them as reverberant if $T_{60}>0.22$~s and $C_{50}<18$~dB.

\begin{figure}[tb]
	\centering
	\includegraphics[width=\columnwidth,clip,trim=175 100 210 100]{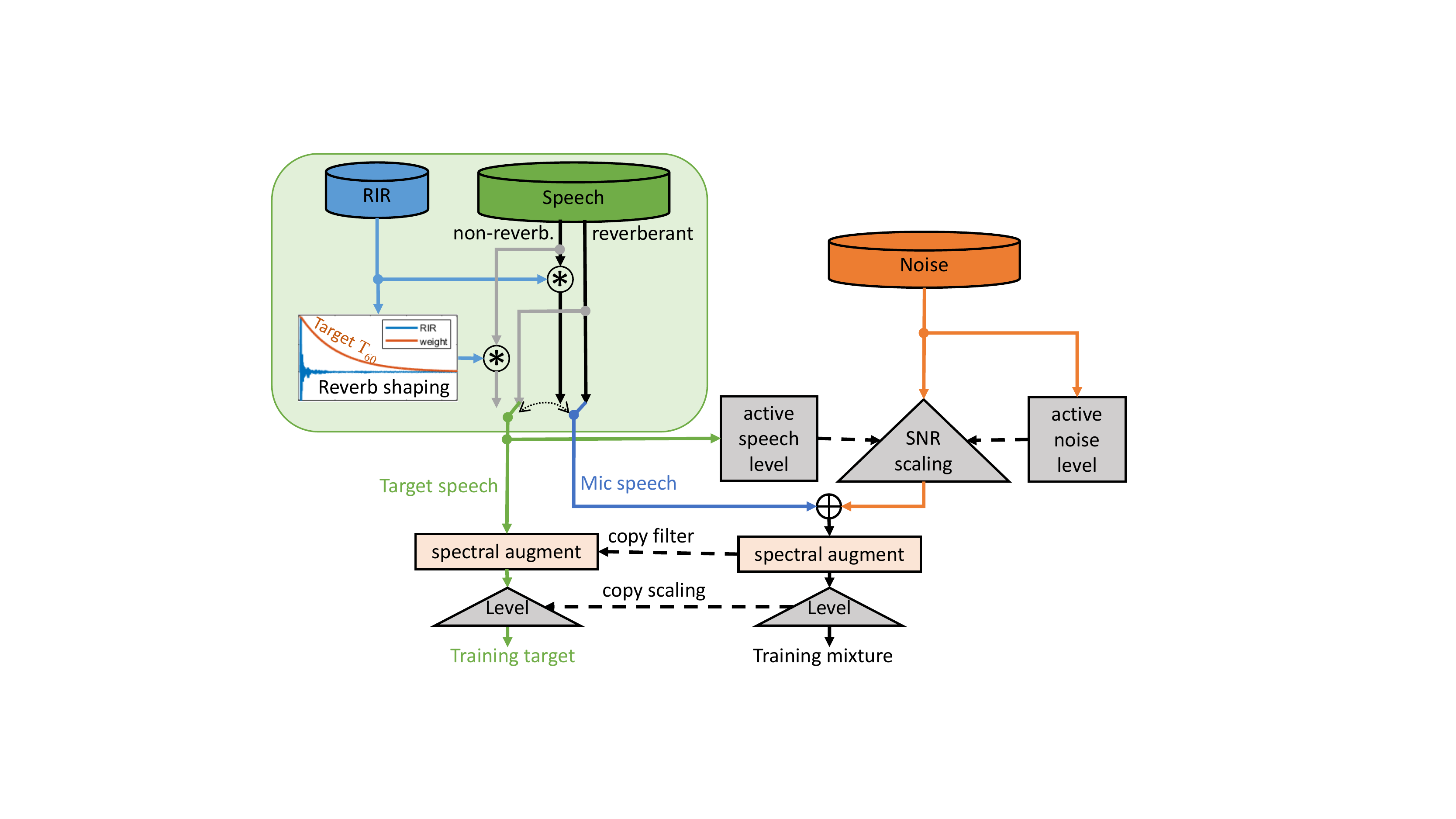}
	\caption{Training data generation: Reverberant speech is used as is, while non-reverberant speech is augmented with \acp{RIR}, and the training targets are created using shaped \acp{RIR}.}
	\label{fig:data_gen}
\end{figure}

Our data generation pipeline, outlined in Fig.~\ref{fig:data_gen}, is described in the following.
While already reverberant speech files are mixed with noise as is, non-reverberant speech files were augmented with acoustic impulse responses randomly drawn from a set of 7000 measured and simulated responses from several public and internal databases. 20\% non-reverberant speech is not reverb augmented to represent conditions such as close-talk microphones or professionally recorded speech. To obtain natural sounding, low-reverberant, and time-aligned target speech signals, the reverberant impulse responses were shaped to a maximum decay of $T_{60}^\text{max} = 0.3$~s as shown in Fig.~\ref{fig:data_gen}. The weighting function (shown as red line in the reverb shaping block) is an exponential decay with the desired reverberation time \cite{Polack1988}, starting at the direct sound $t_0$ of the \ac{RIR}
\begin{equation}
	w_\text{RIR}(t) = 
	\begin{cases} 
		\exp\left( -(t-t_0) \frac{6\log(10)}{ T_{60}^\text{max}} \right), & \quad t \geq t_0 \\
		1, & \quad t < t_0
	\end{cases}
\end{equation}
To generate noisy training data, random speech and noise portions were selected to form training clips of 10~s length. Each speech and noise segment is level normalized and concatenated with other clips, if the duration was too short. The reverberant speech segments are then generated as described before, shown in the green box in Fig.~\ref{fig:data_gen}, providing the reverberant speech and non-reverberant target speech signals. The reverberant speech and noise is mixed with a \ac{SNR} drawn from a Gaussian distribution with $\mathcal{N}(5, 10)$~dB. The resulting mixture signals are re-scaled to levels distributed with $\mathcal{N}(-28, 10)$~dBFS. The speech targets are scaled accordingly with the same factors. The optional spectral augmentation was not used here due to the large amount of raw data. Using this pipeline, we created an augmented dataset of roughly 1000~h.

For training monitoring and hyper-parameter tuning, we generated a synthetic validation set in the same way as above, using speech from the DAPS\footnote{https://ccrma.stanford.edu/~gautham/Site/daps.html} dataset, and \acp{RIR} and noise from the QUT\footnote{https://research.qut.edu.au/saivt/databases/qut-noise-databases-and-protocols/} database.
The final test results are shown on the public development set of the Interspeech 2020 DNS challenge \cite{Reddy2020}, consisting of 300 real recordings and 300 synthetic mixtures from unseen datasets.

\subsection{Evaluation metrics}
Evaluating speech enhancement algorithms is a complex task, which led to the development of various objective metrics, while subjective ratings are still the gold standard. Recently, \acp{DNN} are developed to predict \ac{MOS} \cite{Avila2019,Reddy2021}. While we evaluated most of the proposed models using crowd-sourced ITU P.808 tests, we show only the predicted DNSMOS \cite{Reddy2021} for consistency of presentation and space constraints here. Nevertheless, all rankings and trends were coherent across crowd-sourced \ac{MOS}, DNSMOS and intrusive objective metrics like \ac{PESQ}, \ac{CD}, and \ac{siSDR} on the synthetic validation set.

For all models, we relate their audio quality to an estimate of the computational complexity during inference in terms of \ac{MAC} operations. Note that we count only the operations related to applying the weights and biases, which usually contribute the major computational burden. We do not account for applying activation functions, and also omit feature extraction, \ac{STFT}, and enhancement operations, which are common for all models, and are both negligible compared to the burden of the \ac{DNN} models.

\subsection{Algorithmic parameters}
%We use a sampling frequency of 16~kHz. 
%We investigated different \ac{STFT} parameters always using 50\% overlapping squareroot-Hann windows, but varying window sizes of 20~ms or 32~ms and correspondingly FFT sizes of 320 or 512 points.
We use a sampling frequency of 16~kHz, an \ac{STFT} with 50\% overlapping squareroot-Hann windows of 20~ms length, and a corresponding FFT size of 320 points. The inputs to the networks are 161-dimensional log power spectra.
We parameterize \emph{NSnet2} models denoted by NSnet2-$R$, where $R$ denotes the number of \ac{GRU} nodes per layer.
We parameterize \ac{CRUSE} with different encoder/decoder sizes, starting always with $C_1 \!=\! 16$ channels, and doubling the channels each layer. CRUSE models are denoted by CRUSE$L$-$C_L$-$N$xRNN$P$, where $L$ are the number of encoder/decoder layers, the last encoder layer filters $C_L$ can vary to scale the RNN layer width, $N$ are the number of RNN layers, and $P$ are the number of parallel RNNs. For example, CRUSE4-120-1xGRU4 has 4 encoder/decoder layers with filters 16-32-64-120, and 1 layer of 4 parallel GRUs. Convolution kernels are always (2,3), unless denoted explicitly as 1D convolutions with (1,3) kernels operating only across frequency.

\section{Results}
\label{sec:results}
%
%To investigate the influence of the network size, especially memory capacity of the recurrent layers, we scale the size of the \acp{RNN} by varying the number of \ac{GRU} nodes and compute the resulting inference complexity in terms of \acp{MAC}. For NSnet2, we simply changed the number of hidden units simultaneously for both \ac{GRU} layers. For CRUSE4, we changed the filters $C_4$ of the last encoder layer, which then naturally changes the size of the central \ac{GRU} layer as well. Fig.~\ref{fig:rnnsize_macs} shows the resulting DNSMOS on the test set directly over the MACs for each network architecture. It is interesting how directly the resulting \ac{MOS} scales with the the recurrent memory size and therefore also the complexity.
%%
%\begin{figure}[tb]
%	\includegraphics[width=\columnwidth]{mac_vs_rnnsize.png}
%	\caption{Scaling the computational complexity by changing the number of recurrent nodes.}
%	\label{fig:rnnsize_macs}
%\end{figure}

%
\begin{figure}[tb]
	\includegraphics[width=\columnwidth,clip,trim=5 2 20 13]{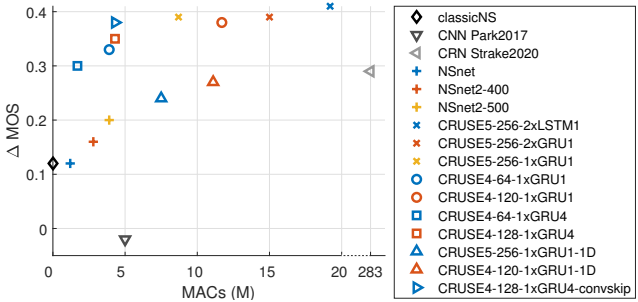}
	\caption{MOS improvement vs. computational complexity (MACs).}
	\label{fig:dnsmos_macs}
\end{figure}
Figure~\ref{fig:dnsmos_macs} shows the tradeoff between computational complexity in terms of \acp{MAC} vs.\ the predicted overall audio quality using DNSMOS \cite{Reddy2021} for several \emph{NSnet2} and \emph{CRUSE} models, and other baselines. In Fig.~~\ref{fig:dnsmos_macs} all CRUSE models use add skips without $1\times1$ conv blocks unless denoted explicitly.
The first baseline is a classic noise suppressor (\emph{classicNS}) exploiting only stationarity of speech \cite{Tashev2009}. While this method non-surprisingly achieves only minor MOS improvement of around 0.12 on the test set including many highly non-stationary noise types, the achievement relative to its computational burden in the order of a few 1000 MACs is a tiny fraction of even our smallest models.
For direct comparison of the following DNN-based baselines, we only took the architectures, but trained them using the same features, prediction targets, loss, and training procedure as for all other networks in this paper: i) \emph{NSnet} \cite{Xia2020} yields similar MOS as the \emph{classic NS}. ii)
The fully convolutional architecture proposed in \cite{Park2017} underperforms in this task, which we mainly suspect to the absence long-term temporal modeling as it uses only 8 frames temporal context. iii) The \emph{CRN-LSTM9} architecture proposed in \cite{Strake2020} yields significant MOS improvement, but is computationally extremely inefficient with 283M MACs due to large CNN filter kernels and wide recurrent layers.

\emph{NSnet2}, shown with different RNN sizes denoted as NSnet2-400 and NSnet2-500, performs better than NSnet and its quality can be improved by increasing the RNN size to 500 units, which also scales up the MACs. The CRUSE5 models gain largely in efficiency by replacing the 2 LSTM layers ({\color{matlab1}$\times$}) with 2 GRUs ({\color{matlab2}$\times$}), and further by going to only a single GRU ({\color{matlab3}$\times$}).
We can observe how moving from fully connected GRUs ($P\!=\!1$) to $P\!=\!4$ parallel GRUs for different RNN sizes ({\color{matlab1}$\bigcirc$}$\rightarrow${\color{matlab1}$\square$} and {\color{matlab3}$\bigcirc$}$\rightarrow${\color{matlab2}$\square$}) reduces complexity with very little performance loss.
The 2D convolutional feature extraction of \emph{CRUSE} is very useful and efficient as moving to 1D convolutions using kernels of (1,3) deteriorates the tradeoff significantly for CRUSE5-256-1GRU ({\color{matlab3}$\times$}$\rightarrow${\color{matlab1}$\bigtriangleup$}) and CRUSE4-120-1GRU ({\color{matlab2}$\bigcirc$}$\rightarrow${\color{matlab2}$\bigtriangleup$}). This seems to contradict findings in \cite{Tan2020b}, where no performance degradation is claimed by using 1D convolutions \emph{on non-reverberant datasets}, which highlights the importance of using appropriate data.

Overall, Figure~\ref{fig:dnsmos_macs} reveals a few very interesting trends: i) The fully recurrent NSnet models have generally lower complexity, but also lower quality than the \emph{CRUSE} models. We can conclude that the convolutional encoder-decoder structures of \emph{CRUSE} are very helpful. Especially using also temporal encoder/decoder layers boosts efficiency.
ii) We can observe a surprisingly linear correlation between MACs and MOS. Consequently for the tested models, the average achieved quality is a monotonically increasing function of the computational effort.
The MAC-MOS linear relation is even more clear for models within the same class, e.g. the NSnet models, or \emph{CRUSE} models with different \ac{RNN} widths.
The most efficient models, i.\,e.\, the proposed \emph{CRUSE} models with parallel \acp{GRU} and optimized skip connections ({\color{matlab1}$\triangleright$}) break out of the linear trend, pushing towards the desired upper left corner.

\begin{figure}[tb]
	\includegraphics[width=\columnwidth,clip,trim=0 2 0 6]{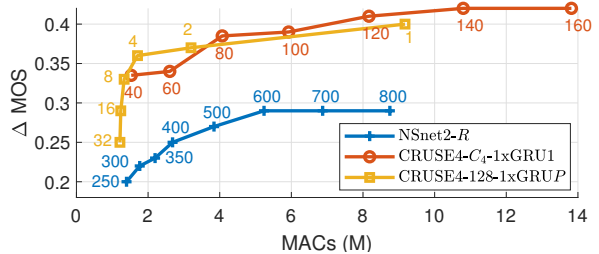}
	\caption{Tradeoff by changing RNN width for NSnet2 and CRUSE, and parallel RNN grouping. Colored numbers denote the variable parameter per model, $R$, $C_4$, and $P$.}
	\label{fig:parallel_grus}
\end{figure}
Fig.~\ref{fig:parallel_grus} illustrates scalable performance depending on the \ac{RNN} width. The blue and red lines show NSnet2-$R$ and CRUSE4-$C_4$-1xGRU1 models, where the width of the RNN was scaled by changing the RNN width $R$ for NSnet2, or changing the last encoder layer's filters $C_4$ and therefore also the RNN width for CRUSE4, respectively. We can observe a clear monotonic relationship between RNN throughput or memory and the obtained quality. Obviously, the complexity vs.\ quality tradeoff deteriorates for too large models. It is a useful property that the performance of these networks can be scaled given a certain computational budget.
The yellow line in Fig.~\ref{fig:parallel_grus} shows the effect of grouping the fully connected RNN for a CRUSE4-128-1xGRU$P$ model into $P$ disconnected RNNs while keeping the RNN feedforward flow fixed. Significantly better tradeoffs can be achieved with $P=2$ and $P=4$.

\begin{table}[tb]
	\centering
	\caption{Effect of skip connection type for CRUSE4-128-1GRU4.}
	\label{tab:cruse_arch}
	\begin{tabular}{l|cc}
		\toprule
		model 					& MACs (M) & $\Delta$MOS \\
		\midrule
		no skips				& 4.3	& 0.32 \\
		add skips	 			& 4.3	& 0.35 \\
		add conv $1\!\!\times\!\!1$ skips	& 4.3	& \textbf{0.38} \\
		concat skips 			& 4.8	& 0.38 \\ %\hline
		%	no padding 				& 3.2	& 3.13 \\
		%	FC merging				& 4.5	& 3.18 \\
		%	depth-sep				& 3.2	& 3.02 \\
		%	convGLU					& 4.4	& 3.14 \\
		\bottomrule
	\end{tabular}
	\vspace{-.1cm}
\end{table}
Table~\ref{tab:cruse_arch} shows further ablations on the skip connections using the CRUSE4-128-1xGRU4P architecture. Addition skips are better than no skip connections. While concat skips improve the MOS, the same MOS can be achieved by inserting cheap $1\times1$ convolutions as shown in Fig.~\ref{fig:skips} by the orange block.

While the execution time of the models is subject to optimization for the targeted hardware platform, we provide a sense of relating the \acp{MAC} to the actual execution time of the efficient CRUSE4-128-1xGRU4 model measured on a Intel\textsuperscript{\textcopyright} Core\texttrademark i7 QuadCore at 3.5~GHz: Without further optimization, the ONNX model processes one audio frame on average in 0.3~ms, resulting in a reasonable CPU utilization of less than 3\% within the hop size budget of 10~ms.
NSnet2-500 runs in 0.15~ms.

\section{Conclusions}
We proposed a flexible and scalable \ac{CRN} for speech enhancement, trained on large data and tested on real recordings. We show that using simple spectral suppression based networks of comparatively small size can achieve substantial quality improvement when trained on a representative dataset with a suitable loss taking the time domain reconstruction into account. We show that the obtained speech quality is a function of network size, especially depending on the recurrent layer width. We show gains on the speech quality vs. computational complexity tradeoff by modified skip connections and a disconnected parallel RNN structure. While the proposed models use only a fraction of the computational budget of standard CPUs in real-time, the quality gain per computational burden compared to a traditional speech enhancement method is still less efficient and can hopefully be further improved in the future.
	
	%\vfill\pagebreak
	\balance
	
	\bibliographystyle{IEEEbib}
	\bibliography{sapref.bib}
	
\end{document}